\documentclass[aps,prl,showpacs,floats,twocolumn,floats,superscriptaddress,floatfix]{revtex4}

\usepackage{graphicx}
\usepackage{bm}
\usepackage{amsfonts}
\usepackage{color}

\usepackage{amsmath}    
\usepackage{epsfig}
\usepackage{subfigure}  
\usepackage{hyperref}   
\usepackage{bm}
\usepackage{amssymb}

\def\bu{{\boldsymbol u}}
\def\buh{{\hat {\boldsymbol u}}}

\newcommand{\be}{\begin{equation}}

\newcommand{\ee}{\end{equation}}

\newcommand{\bk}{{\boldsymbol k}}

\newcommand{\bx}{{\boldsymbol x}}
\newcommand{\br}{{\boldsymbol r}}

\newcommand{\bF}{{\textbf F}}
\newcommand{\bfc}{{\boldsymbol f}_c}
\newcommand{\bn}{{\boldsymbol \nabla}}

\newcommand{\romeaddress}{Department of Physics and INFN, University
  of Rome ``Tor Vergata'', Via della Ricerca Scientifica 1, 00133,
  Rome, Italy.}

\newcommand{\tueaddress}{Department of Applied Physics, Eindhoven
  University of Technology, 5600 MB Eindhoven, The Netherlands and
  Istituto per le Applicazioni del Calcolo, Consiglio Nazionale delle
  Ricerche, 00185 Rome, Italy.}

\begin{document}

\title{The statistical properties of turbulence in the presence of a smart small-scale control}

\author{Michele Buzzicotti} 
\affiliation{\romeaddress}
\author{Luca Biferale}
\affiliation{\romeaddress}
\author{Federico Toschi} 
\affiliation{\tueaddress}
\email{f.toschi@tue.nl}

\begin{abstract}
  By means of high-resolution numerical simulations, we compare the
  statistical properties of homogeneous and isotropic turbulence to
  those of the Navier-Stokes equation where small-scale vortex
  filaments are strongly depleted, thanks to a non-linear extra
  viscosity acting preferentially on high vorticity regions. We show
  that the presence of such {\it smart} small-scale drag can strongly
  reduce intermittency and non-Gaussian fluctuations.  Our results
  pave the way towards a deeper understanding on the fundamental role
  of degrees of freedom in turbulence as well as on the impact of
  (pseudo)coherent structures on the statistical small-scale
  properties. Our work can be seen as a first attempt to develop
  smart-Lagrangian forcing/drag mechanisms to control turbulence.
\end{abstract}
\maketitle

\section{Introduction}
Fluid dynamics turbulence is characterized by intermittent and
non-Gaussian fluctuations distributed over a wide range of space- and
time-scales \cite{Frisch95,benzi2010inertial,yeung2015extreme,iyer2017reynolds,sinhuber2017,popebook}. In the limit of
infinite Reynolds numbers, $Re$, the number of dynamical degrees of
freedom tends towards infinity, $\#_{dof} \sim Re^{9/4}$, where $Re =
U_0L_0/\nu$ with $\nu$ the viscosity,  $U_0$ and $L_0$ the typical velocity and large-scale
in the flow, respectively. Are all these degrees of freedom equally
relevant for the dynamics? Do extreme events depend only on some
large-scale flow realizations? Can we selectively control some
degrees-of-freedom by applying an active forcing and/or drag? These
are key questions that we start to answer by using high resolution
numerical studies of the three dimensional Navier-Stokes
equations. The long term goal is twofold. {First, we are interested to have a new numerical tool to
  ask novel questions concerning the statistical and topological properties of specific flow structures. Second,  we aim
  to develop useful (optimal) control
strategies to suggest forcing protocols that may
be implemented in laboratory experiments, where the flow can be seeded
with millions of passive or active particles, preferentially tracking
special flow regions \cite{bec2007heavy,toschi2009lagrangian,calzavarinijfm2009,qureshi2008,mercado2010,gibert2012,gustavsson2016,mathai2016}}. For
example, we nowadays know how to actively control spinning properties
of small magnetic particles \cite{Stanway04,Falcon17}, how to blow-up small bubbles
by sound emissions \cite{abe02,hauptmann13,hauptmann14} and/or how to assemble micro-metric
objects with a self-adaptive shape depending on the flow rheological
properties \cite{huang2019}. Recent developments in 3d-printing and
micro-engineering technologies promise that new tools will be
available in the next few years for fluid control or fluid
measurements in the laboratory. We believe that these new tools could
be capable to do, in a ``smart'' way, what dummy and passive polymers
already do in controlling drag and flow correlations \cite{lumley73drag,white2008mechanics,Fischer11}.  In
this paper, we perform a first attempt to modify/control fluid
turbulence by adding a small-scale forcing only on intense vorticity
regions. We start from the case where the forcing is always
detrimental, i.e. removes energy. The idea is to have a numerical  experiment mimicking the effects of small-particles that
preferentially track high vorticity regions (i.e. light bubbles) and
that can be activated such as to spin or blow-up and increase
the drag locally. This is only one potential protocol over a wide and
broad range of other applications to many others flow conditions at
high and low Reynolds.\\
\begin{figure}[htp]
\centering
\includegraphics[scale=0.25]{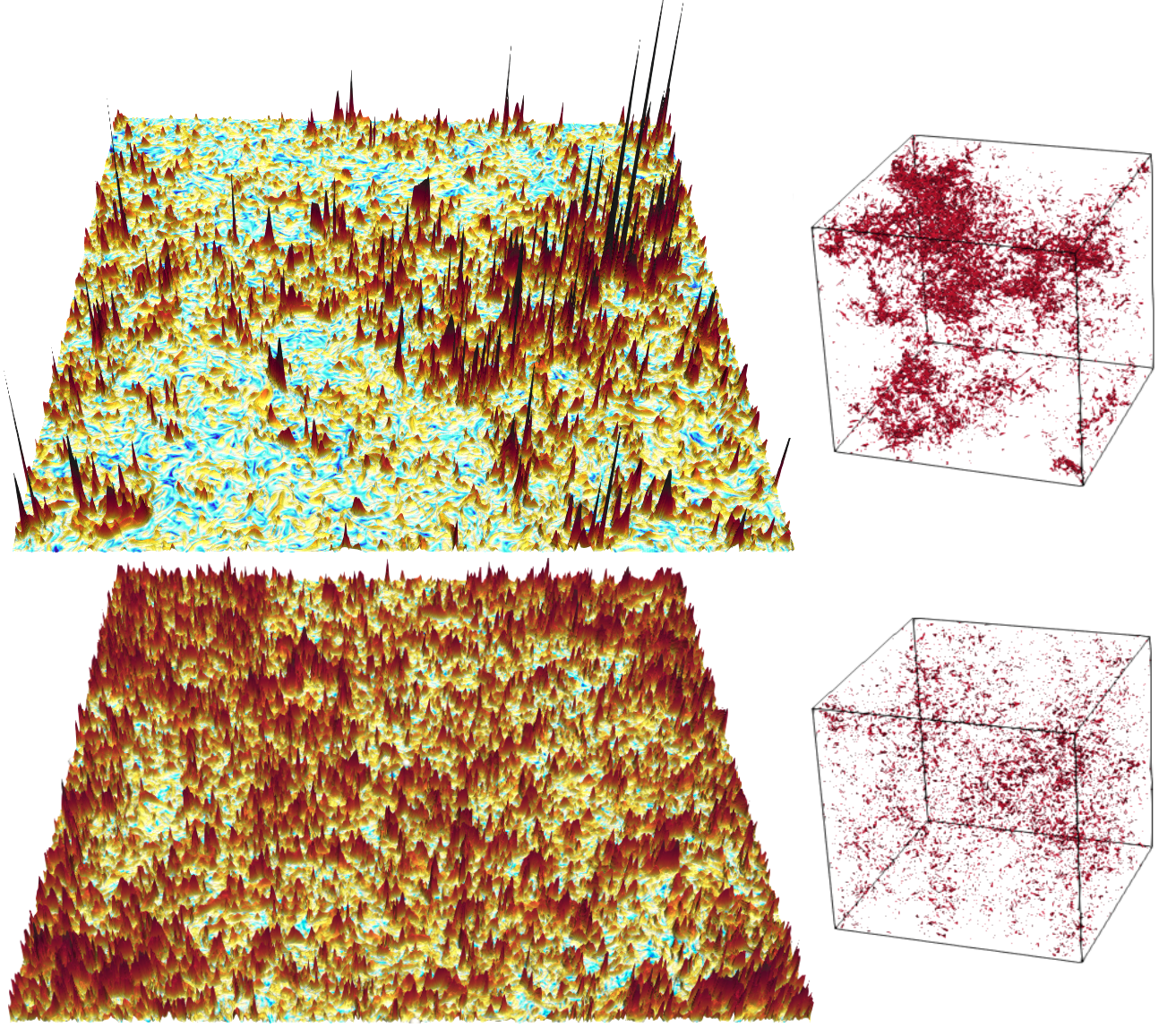}
\caption{(Top row) Left: visualization of vorticity amplitude in a 2d plane,
  from a simulation of NSE without control term ($\beta=0$). Right:
  isocontour regions where the vorticity amplitude is above $20\%$ of its maximum value measured over the flow
  volume. (Bottom row) Same visualizations obtained from a simulation
  where the control term is acting on the dynamics of the NSE. Left:
  visualization of enstrophy plane. Right: contour regions where the
  vorticity amplitude is above the control forcing threshold fixed at
  $\omega_p=0.2\omega_{max}$. The control forcing amplitude used in
  the simulation presented here is $\beta=5$. Both simulations are
  performed with a number of collocation points $N=1024^3$.}
\label{fig:visual}
\end{figure}
{\sc Method.} We consider the Navier Stokes equations (NSE) for an
incompressible flow, subjected to two different types of forcing
mechanisms: \be
\label{eq:navierstokes}
\partial_t \, \bu + \bu \cdot \bn \bu = - \bn P + \nu \Delta \bu + \bF
- \bfc \ee where $\bF$ is a standard large-scale stirring mechanism
while $\bfc$ is a second forcing which acts --in our implementation--
as a control term on the small-scales dynamics. In particular, in this
paper, we will only consider an external smart-drag, proportional to
the velocity $ \bfc(\bx,t) = c(\bx,t) \bu(\bx,t) $ and acting such as
to preferentially depleting only those regions where vorticity is
important \be
\label{eq:forc_om}
c(\bx,t) = \beta \, \left ( \frac{\text{tanh} \left[
    (\omega(\bx,t)-\omega_{p})\right ]+1}{2} \right ),
\ee 

where $\omega(\bx,t) = |\bn \times \bu|$, is the vorticity
intensity, $\omega_p$ is one threshold above which the control term is
strongly active and $\beta$ is an overall rescaling factor of the
control amplitude, hence $\beta=0$ would correspond to the usual NSE
without control. From its definition it is possible to see that
$\bfc(\bx,t)$ is always close to zero except inside structures
dominated by the intense vortex filaments, where the $\tanh$ become
positive and equal to 1.  The region where $\bfc(\bx,t)$ is acting can
be tuned by changing the threshold $\omega_p$, whose value has been
fixed as a percentage of the maximum vorticity, $\omega_{max}$,
measured in the stationary state of a simulation without the control
term, hence: 

$$\omega_p = p \, \omega_{max}$$ 

with $0< p \le 1$.  In the transition
region around the isoline where $\omega(\bx,t)=\omega_{p}$ the control
function (\ref{eq:forc_om}) will introduce compressibile effects in
(\ref{eq:navierstokes}).  Therefore, before adding the control term
to (\ref{eq:navierstokes}) one needs to project it on its solenoidal
component.

\begin{figure}[htp]
\centering
\includegraphics[scale=0.5]{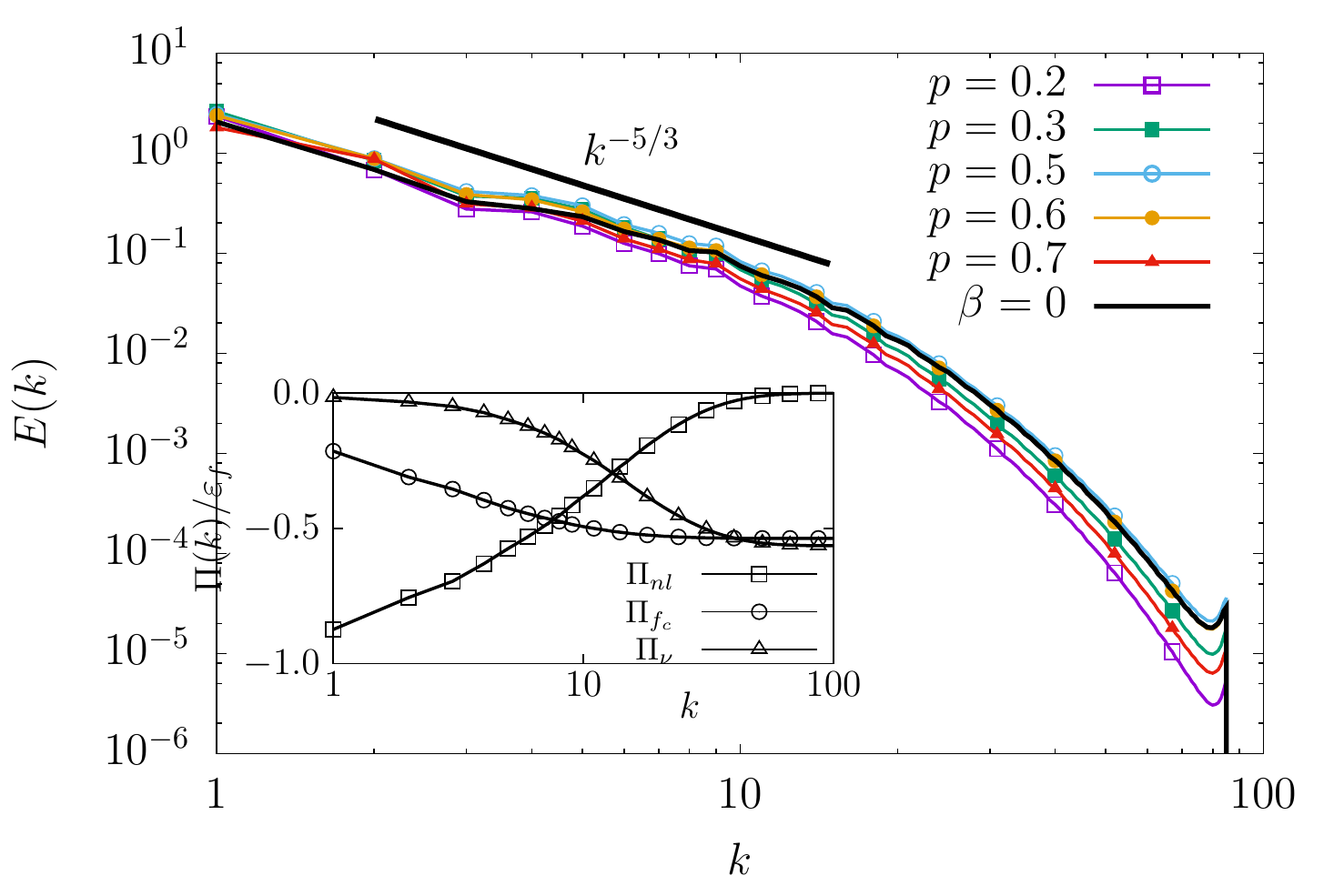}

\caption{Main panel: energy spectra averaged on time for different
  simulations at changing {the control} threshold,
  $\omega_c/\omega_{max} = p$, with a fixed amplitude,
  $\beta=5$. Notice that $\beta=0$ corresponds to the uncontrooled full NSE. (Inset)
Energy fluxes from the non-linear term, $\Pi_{nl}$, from the control
term, $\Pi_{f_c}$, and from the viscous dissipation, $\Pi_{\nu}$,
normalized to the mean energy input, $\varepsilon_f$. Here  $p=0.2$ and $\beta=5$.
}
\label{fig:energy}
\end{figure}


The projection operation breaks the
local positive definiteness of the control term, which remains purely
dissipative only globally as an average on the whole volume.

{\sc Results.} In Fig.~\ref{fig:visual} we present two visualizations
of a plane of the vorticity intensity in the stationary state for two
simulations, one for the standard NSE (top panel) and one with the
control term acting on the flow (bottom panel). The two planes in
Fig.~\ref{fig:visual} are warped upwards depending on the vorticity
values, in this way it is possible to see that the intense peaks
developed by the NSE are pruned by the small-scales forcing in the
controlled dynamics. From the figure it is qualitatively evident that
vorticity is strongly depleted when the small-scale drag is acting, as
expected. In the same figure, next to the vorticity planes, we show a
3D rendering of the contour regions where the vorticity value is above
$20\%$ of its maximum value, for the case of the uncontrolled NSE (top
panel) and the contour regions where the vorticity value is above the
forcing threshold, $p=0.2$, for the case of the controlled flow
(bottom panel). From the volume rendering we can appreciate that the
control forcing tends to homogenise the spatial distribution of the
intense vorticity events while they result more intermittent and
localized when the dynamics is not controlled. It is also interesting
to observe  that the volume fraction where the forcing is
acting is very small even though in those visualizations we are using
a broad threshold in terms of the vorticity values,
$p=\omega_p/\omega_{max}=0.2$.

{\sc Energy balance.}  As already mentioned, the control term has a
 dissipative global effect on the turbulent dynamics which goes in addition to
the normal dissipation produced by the kinematic viscosity. In this
way, a second possible channel is opened where the energy, injected by
the large scales forcing, can be dissipated.  The total energy balance
equations becomes: \be
\label{eq:energybal}
\frac{1}{2}\partial_t \langle \bu^2 \rangle  = \nu \langle \Delta \bu^2 \rangle - \langle \bfc \cdot \bu \rangle + \langle \bu \cdot \bF \rangle.\\
\ee
wehere we have the total kinetic energy, $E = \frac{1}{2}
\langle \bu^2 \rangle$, the viscous dissipation $\varepsilon_{_\nu} =
\nu \langle \Delta \bu^2 \rangle$, the dissipation induced by the
control mechanism, $\varepsilon_c =\langle \bfc \cdot \bu \rangle$, and
the energy injection rate $\varepsilon_{_{f}} = \langle \bF \cdot \bu
\rangle$, and  with $\langle \bullet \rangle$ we intend an average on
the whole volume. \\
{ \sc Numerical Simulations.} To assess the statistical properties of
Eq.~\eqref{eq:navierstokes} a set of direct numerical simulation have
been performed at changing resolution and the control parameters,
namely $\beta$ and $\omega_p$.  We used a pseudo-spectral code  with
resolutions up to $1024^3$ collocation points in a triply periodic
domain $\Omega$ of size $L = 2\pi$. Full $2/3$-rule de-aliasing is
implemented (see Table I for details). The homogeneous and isotropic external force, $\bF$, is
defined via a second-order Ornstein-Uhlenbeck process \cite{biferale2016coherent}.  All simulations where  control is on, have been
produced starting from a stationary configuration of the uncontrolled
case $\beta=0$ and all statistical quantities are calculated after
that a new stationary state is achieved.

\begin{table}
    \begin{center}
    \begin{tabular}{|c|ccccc|}
    Control & $N$ & $\beta$ & $p$ & $\varepsilon_f$ & $\nu$\\
    \hline
    Off & $256$  & - & - &  2.2 & $5.2 \times 10^{-3}$\\
    Off & $1024$ & - & - &  5.5 & $8 \times 10^{-4}$\\
    \hline
    On  & $256$  & $[0.1 \div 50]$ & $[0.1 \div 0.7]$  &  2.2 & $5.2 \times 10^{-3}$\\
    On  & $1024$ & $50$            & $[0.05 \div 0.6]$ &  5.5 & $8 \times 10^{-4}$\\
    \end{tabular}
    \end{center}
    \caption{Parameters used in the  simulations.  Control:
      indicates if the control term ($-c(\bx,t)\bu(\bx,t)$) is applied
      (On) or not (Off) in Eqs.~\eqref{eq:navierstokes}; $N$ is the number
      of collocation points in each spatial direction; $\beta$ is the
      amplitude of the control term; $p$ is the percentage of the maximum vorticity above
      which the control term is active $\omega_p=p \omega_{max}$;
      $\varepsilon_f$ is the mean energy input injected by the large scales
      forcing; $\nu$ is the kinematic viscosity. The amplitude of
      Ornstein-Uhlenbeck forcing is  $f_0=0.16$ and $f_0=0.14$
      for  $N=256$ and $N=1024$ respectively; the
      correlation time is  $\tau_f=0.6$ for      $N=256$ and $\tau_f=0.23$ for $N=1024$. The forcing is
      active on the  window $k_f=[0.5:1.5]$ for resolution $N=256$ and on  $k_f=[0.5:2.5]$ for $N=1024$. The
      Kolmogorovo scale is  $\eta=
      (\nu^3/\varepsilon)^{1/4}$, where $\varepsilon$ is the
      dissipation rate. Resolution is kept at $\eta/dx \ge 0.7$.}
    \label{tbl:simulations}
\end{table}
In Fig.~\ref{fig:energy}  we present the time average of the instantaneous  energy spectra:
\be
\label{eq:spectrum}
E(k,t) = 0.5 \sum_{k< |\bk|<k+1} |\buh(\bk,t)|^2 \ee which are almost
independent of the control parameter, $p$.  Only for the smallest
value of $\omega_p$, with $p \sim 0.2$, we can notice a small energy
depletion at large wavenumbers. However, in all cases, the inertial
range scaling properties are unchanged with the slope very close to
the Kolmogorov's prediction $k^{-5/3}$.  In the inset of the same
figure we show for the controlled simulation with $p=0.2$ and
$\beta=5$, the balance of the energy flux produced by the non-linear
term, $\Pi_{nl}(k)$, by the viscous drag, $\Pi_{\nu}(k)$ and by the
control forcing, $\Pi_{f_c}(k)$. In the stationary state we can write
the Fourier space energy balance equation as; \be \Pi_{nl}(k) +
\Pi_{f_c}(k)+ \Pi_{\nu}(k) = \varepsilon_f \, ,
\label{eq:en_bal}
\ee
where $\varepsilon_f$ is the large scales energy input of the stochastic forcing. From the inset of Fig.~\ref{fig:energy} we can see that the control forcing is mainly active in the high wavenumbers where its contribution equals the one from the viscous dissipation, while at small/intermediate wavenumbers the non-linear interactions remain the leading one.



\begin{figure}[htp]
\centering
\includegraphics[scale=0.55]{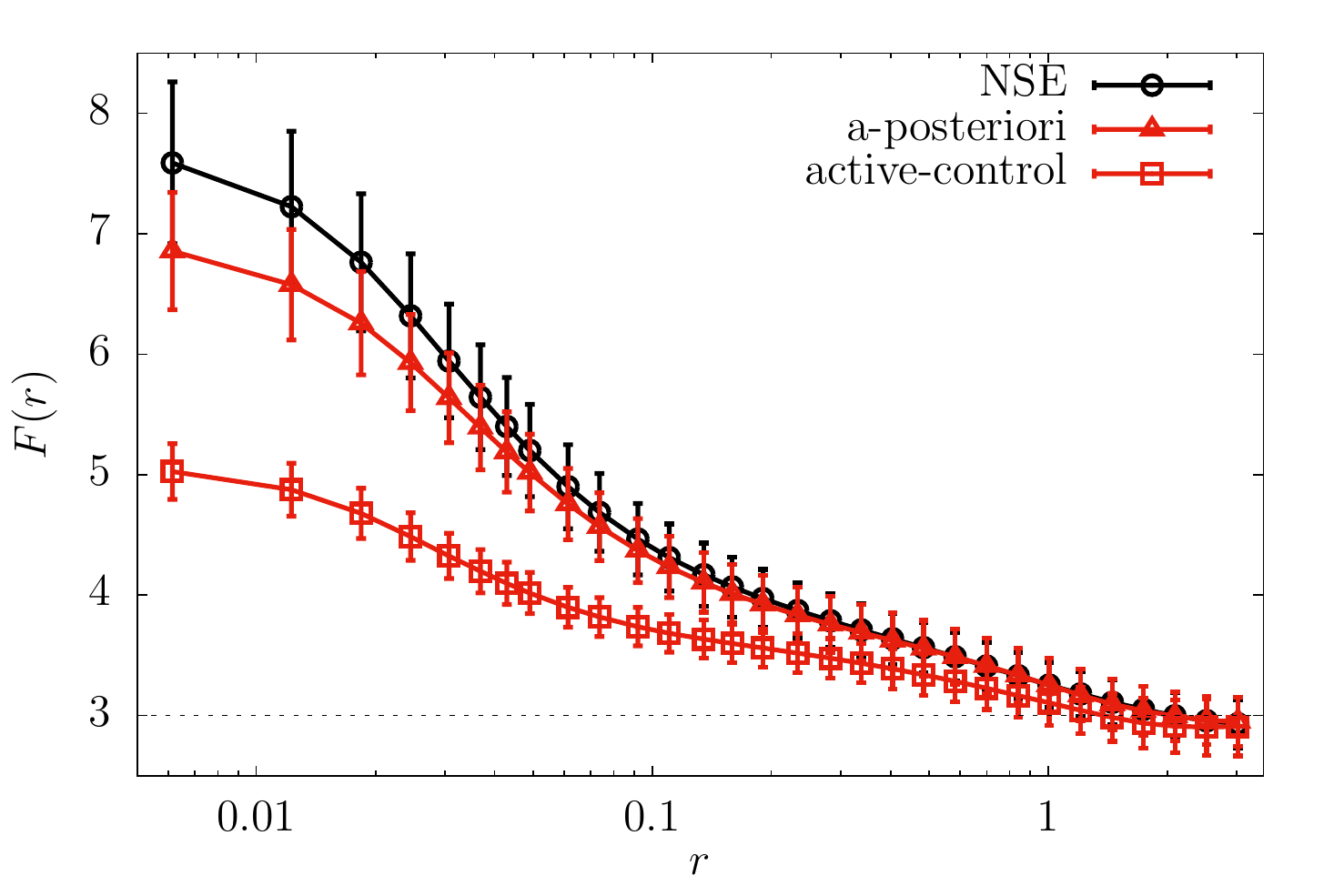}
\caption{Log-log plot of the flatness, $F(r)$, versus
  $r$ measured from $30$ different snapshots in time during the
  evolution of a system with size $N=1024^3$. The black lines with
  open circles (NSE) are data from a simulation without control term ($\beta=0$),
  while the red line with open squares (active-control) are data from
  the controlled NSE, using a forcing threshold
  $\omega_c/\omega_{max}=0.2$ and an amplitude $\beta=50$. The black
  line with full circles (NSE High-Visc) is, again, the NSE without
  control term but with a higher viscosity value in a way to have the
  same total drag coefficient, $d_{tot}$, of the controlled
  simulation. The last curve shown, red line with empty triangles
  (a-posteriori), is the flatness measured from the same simulation
  without dynamical control but skipping from the average all regions
  in the volume where the vorticity module is above the forcing
  threshold $\omega_p/\omega_{max}=0.2$. In all curves, errors are
  evaluated as the standard deviation from $30$ configurations.}
\label{fig:FIG4}
\end{figure}

\begin{figure}[htp]
\centering
\includegraphics[scale=0.50]{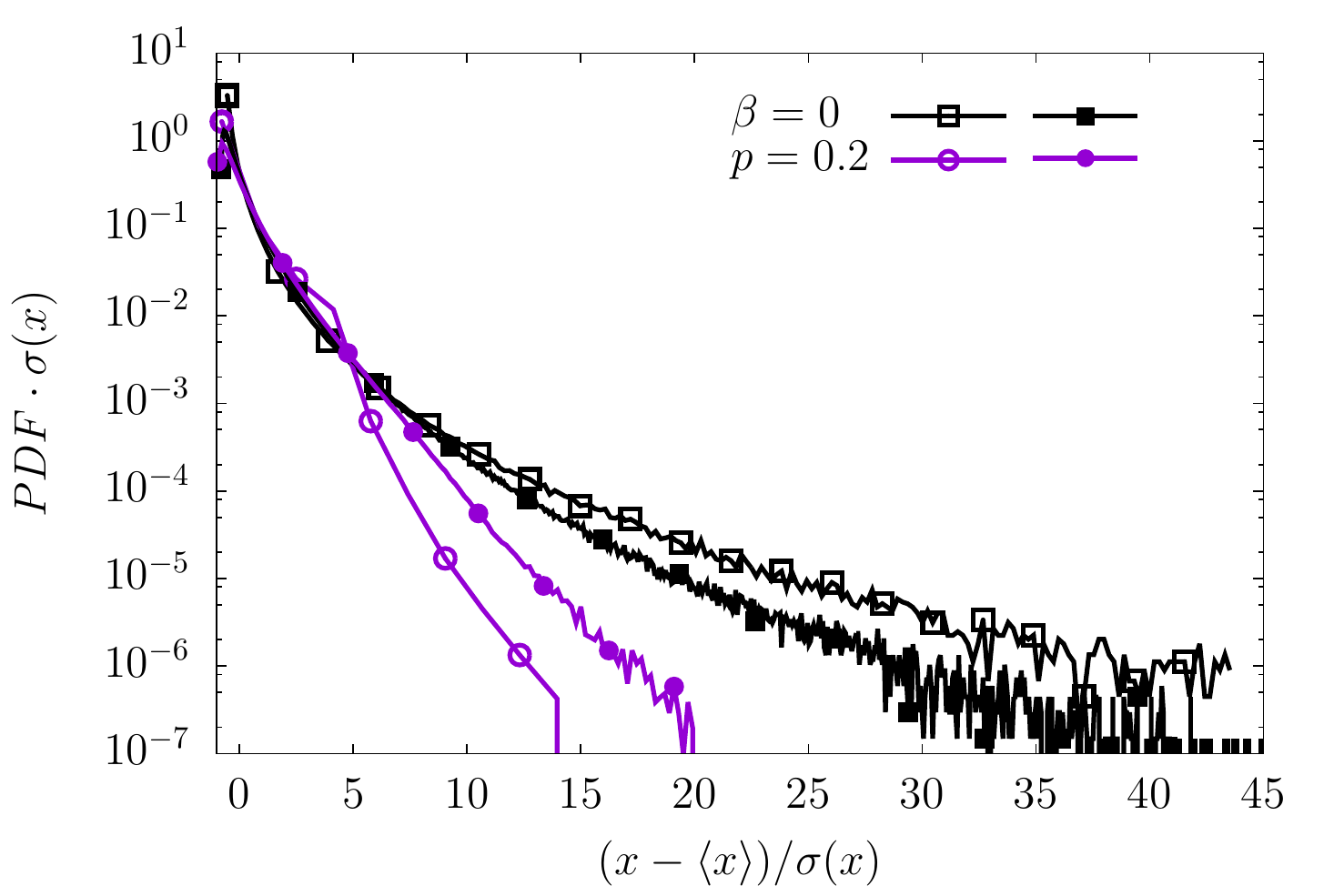}
\caption{ Comparison of the PDFs of enstrophy, $|\bn \times \bu|^2$,
  (open symbols) and shear intensity, ${\cal S}$ 
   (full symbols)
  measured from simulations of standard NSE (black line) and from the
  system controlled with forcing threshold,
  $\omega_p/\omega_{max}=0.2$ and amplitude $\beta=5$.}
\label{fig:FIG2}
\end{figure}

\begin{figure}[htp]
\centering
\includegraphics[scale=0.55]{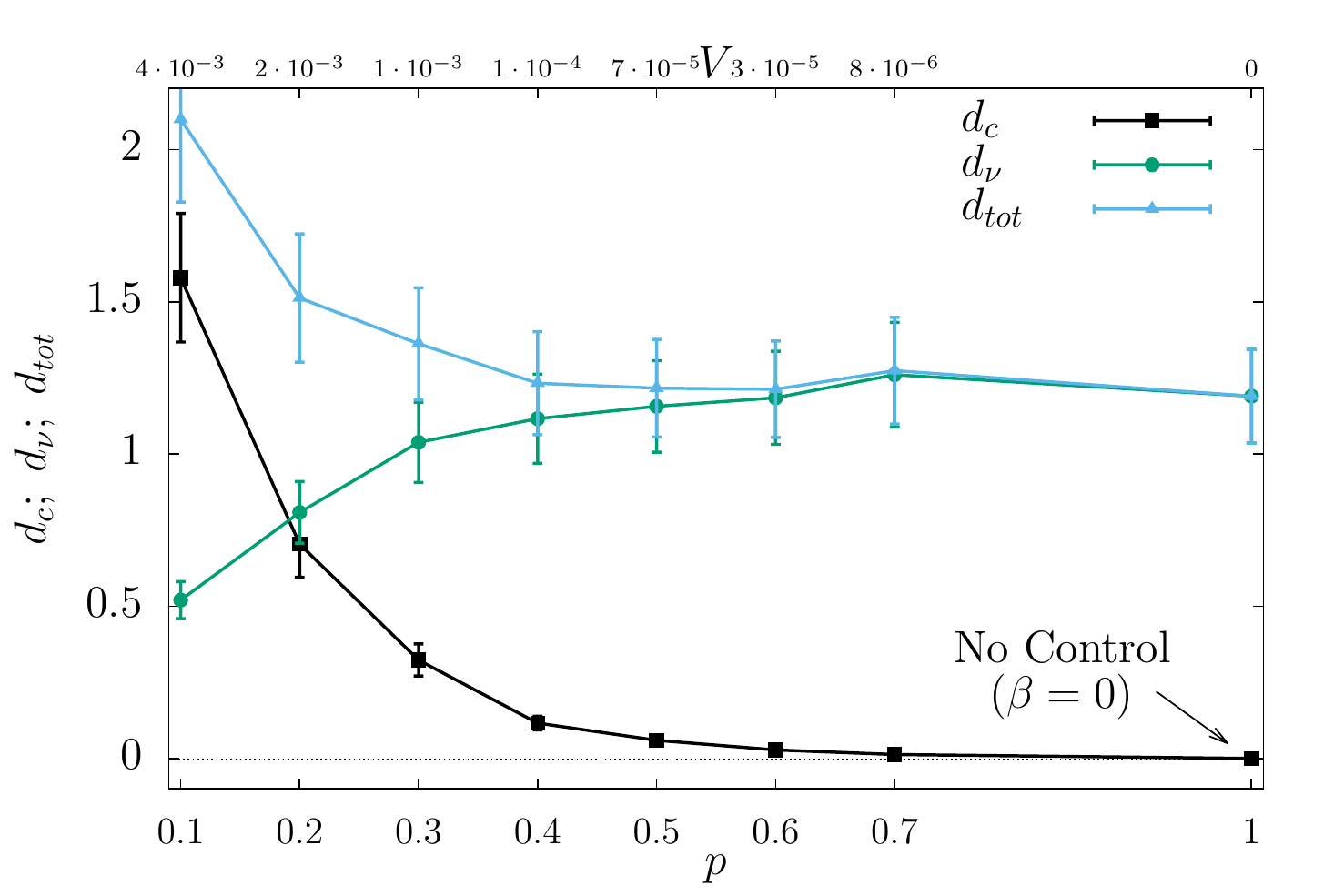}
\caption{Drag coefficient for the viscous dissipation $d_\nu$ (green
  line), the control forcing $d_c$ (black line), and their sum
  $d_{tot}$ (cyan line). Results are shown as a function of the
  threshold, $\omega_p/\omega_{max}=p$, and the volume fraction, $V$, (upper scale)
  where the control forcing is acting. Data are measured from
  simulations with $N=256^3$ and a control forcing amplitude
  $\beta=5$. Errors are evaluated as the standard deviation of the
  temporal fluctuations observed for the different quantities.}
\label{fig:drag}
\end{figure}

{\sc Configuration space statistics.} In the following we analyse the
statistics of the longitudinal velocity increments defined as
$\delta_r u = (\bu(\bx+\br)-\bu(\bx))\cdot \br/r$. In particular we
are interested in the assessment of the effects produced by the
control term on the intermittent properties of the NSE. To do that we
study the scaling properties of the longitudinal structure functions
(SF)   defined as:
\begin{equation}
S_p(r) \equiv \langle [\delta_r u]^p\rangle \sim r^{\zeta_p}.
\label{eq:SFlong}
\end{equation}
Intermittency is measured by the departure of the scaling exponents
from the Kolmogorov 1941 prediction, $\zeta_p=p/3$ in the inertial range,  $\eta < r <L_0$.  In particular,
any systematic non-linear dependency on the order of the moment will
induce a scale-dependency in the flatness, defined by the
dimensionless ratio among fourth and second order SF:
\begin{equation}
F(r) =\frac{S_4(r)}{\left[S_2(r)\right]^2}.
\end{equation}
The flatness for the controlled turbulent flow at resolution
$N=1024^3$ is presented in Fig.~\ref{fig:FIG4}, for the case with
$p=0.2$, compared with the uncontrolled case {$\beta=0$} and with the
uncontrolled case but with an {\it a-posteriori} pruning of all events
where $\omega > \omega_c$. The latter measurement is introduced in
order to understand how much the dynamical pruning imposed by the
evolution of eqn. (\ref{eq:navierstokes}) is different from a simple
conditioning on small-vorticity events taken on the full uncontrolled
NSE. As one can see comparing the empty circles (full {$\beta=0$} NSE) with
the empty squares (active control with $p=0.2$) the effects on the
flatness are dramatic, with both a $100\%$ reduction on the smallest
scale and a decrease of the scaling slope in the inertial
range. Similarly, by comparing the results with the a-posteriori
conditioning (empty triangles) we see that indeed it is crucial to have a
dynamical control to deplete intermittency. To our knowledge {this is
the first evidence that intermittency can be strongly depleted in a
dynamical way with a dynamical criterion based on configuration-space filtering,} at difference from what
obtained by fractal pruning in \cite{lanotte15,buzz16burg,lanotte16,buzz16lag}.
In Fig. ~\ref{fig:FIG2} we show the effects of the vorticity control
point-by-point in the flow volume, by plotting the standardised
probability density function (PDF) for the instantaneous and local
enstrophy, $|\nabla \times \bu|^2$, and shear intensity, 
${\cal S} = \sum_{ij} (\partial_i u_j +\partial_j u_i)^2 $, for one case of active control, $p=0.2$, and compared with
the no-control,{$\beta=0$} case. There are two interesting things to
remark. First, when the control is active, the far tails of the
vorticity are markedly depleted, with almost a sharp cut-off at
$\omega \sim \omega_c$, which is the clear signature that the control
is able to deplete intense vorticity events and to not allow them to
grow again during the evolution. This fact is also good news from a
sort of min-max approach, it means that the amount of control needed
is not too high, being very efficient in stopping the formation of
strong vorticity.  The second interesting point to remark is that the
preferential depletion on vorticity is indeed changing the topological
distribution of extreme events in the flow: from the standard case
where they are mainly given by high vorticity where no control exist
to the case where the extreme fluctuations (far right tails) are more
dominated by strong shear events.\\
    {\sc Drag reduction.} Going back to the observation of the mean
    quantities, it is interesting to estimate the effect of the smart
    forcing on the drag coefficient of the system. Indeed the new
    smart-control allows the system to preferentially dissipate energy
    inside the vortical regions where it is active. To quantify its
    effect {we go back to the balance (\ref{eq:energybal})}  and  split the total drag, $d_{tot}$, in two contributions,
    $d_\nu$ and $d_c$ as follows: \be d_{tot}=d_\nu+d_c, \qquad
    d_\nu=\frac{\varepsilon_{\nu}\,\, L_0}{u_{rms}^3}; \qquad
    d_c=\frac{\varepsilon_c\,\, L_0}{u_{rms}^3}
\label{eq:drag_coeff}
\ee In Fig.~\ref{fig:drag} we show the mean drag coefficients as a
function of the vorticity threshold $\omega_p = p \, \, \omega_{max}$
for the simulations with $N=256^3$ collocation points and with a
moderate control amplitude, $\beta=5$. Fig.~\ref{fig:drag} shows that
the drag contribution coming from the control term is negligible up to
a threshold $p\sim 0.6$, instead moving towards lower thresholds the
dissipation produced by the small-scales term increases and, around
$p=0.2$, the kinematic viscosity and the control dissipations become of
the same order. Moving further the threshold towards lower vorticity
values the control term becomes the leading contribution responsible
for the energy dissipation. In this way, a drag enhancement is
observed for the smaller threshold value and the overall drag
coefficient is increased almost by a factor $2$ compared to the free
NSE at $p=1$.\\
    {\sc Conclusions.}
 { 
    We have presented a first implementation of a smart small-scale control scheme for turbulent flows, based on preferentially damping high vorticity regions.
    In this study, we have shown that the extra drag exerted on the vortex filaments produce a strong reduction on configuration-based intermittency, with depletion of fat tails and rare events in the vorticity field. The topological relative weight of rotational and extensional regions is also affected abruptly. The overall damping of vortex filaments leads to a sort of drag increase. This study open the way to explore other control Lagrangian  mechanism, e.g. based on the heavy-light particles preferential concentration and/or other smart-particles that can be self activated  or activated by external control fields, as for the case of magnetic objects. Optimisation of the particles' properties to track specific flow region can also be attempted in order to enhance/deplete only specific fluctuations \cite{reddy16,colabrese17flow,colabrese18smart,waldock18,novati18}.}

\bibliographystyle{apsrev4-1} 

%
\newpage

\end{document}